\documentclass[showpacs,showkeys,amsmath,amssymb]{revtex4}

 \newcommand{\be}{\begin{equation}}
 \newcommand{\ee}{\end{equation}}

\begin{document}
\title[d-Dimensional Isotropic Harmonic Oscillator]
{Factorization Method for d-Dimensional Isotropic Harmonic Oscillator and the Generalized  Laguerre Polynomials}
\author{Metin Ar\i k\footnote{email:metin.arik@boun.edu.tr}}
\affiliation{Bo\~gazi\c{c}i University,  Faculty of Arts and Sciences, Department of Physics,
34342, Bebek, Istanbul, Turkey}
\author{Melek Baykal, Ahmet Baykal\footnote{email:abaykal@nigde.edu.tr}}
\affiliation{Ni\~{g}de University, Faculty of Science and Letters, Department of Physics, 51240,  Ni\~{g}de,  Turkey}

\begin{abstract}
The factorization method of Infeld and Hull is applied to
the radial Schr\"{o}dinger equation for $d$-dimensional
isotropic harmonic oscillator and various new ladder operators are defined.
The radial energy eigenstates are expressed in terms of the generalized Laguerre polynomials
and their properties are
shown to follow from the expressions  involving  the ladder operators.
In the same way as the harmonic oscillator we also
obtain the bound energy eigenstates
of the Morse oscillator.
\end{abstract}

 \maketitle

 \section{Introduction }

The factorization method for solving the eigenvalue equations for the Schr\"{o}dinger operators
was first introduced by Schr\"{o}dinger  \cite{sch1},
which he also applied it  to the hypergeometric differential equation \cite{sch2}.
Later, the method was generalized by  Infeld and Hull \cite{Infeld-Hull},
and an  extensive list of references  of the method can be found in
\cite{dong}. It is an  effective algebraic method
which is related to the mathematical structure of
supersymmetric quantum mechanics \cite{witten,haymaker,cooper} and
also to the concept of the shape invariance \cite{dutt, gendenshtein}.

The linear harmonic oscillator is the   textbook-example for which  the  method is applied elegantly. 
In this case, all the analytical properties of the eigenfunctions,
namely, the  Hermite polynomials, can be derived from the factorization method
itself. However, this is not the case for  the multi-dimensional isotropic
harmonic oscillator for which the special function related to the radial eigenfunctions 
is the generalized Laguerre polynomials. The generalized Laguerre polynomials occur in
various other well-known quantum mechanical problems, such as
the Morse  oscillator and  the Coulomb problem
and thus their properties are essential for the study of these problems as well.

One of the algebraic methods  to study the spectrum of the isotropic   harmonic oscillator
problem  is to employ a realization of  $\mathfrak{so}(2,1)$
Lie algebra as a spectrum generating algebra \cite{cook}.
Another algebraic method is to transform the radial equation into a confluent hypergeometric equation
and then factorize the resulting equation which requires the use of the properties of the generalized 
Laguerre polynomials \cite{Infeld-Hull,dong}. Still another algebraic way is to start from the generalized 
Rodrigues formula for the Laguerre  polynomials and define ladder operators from various recurrence relations 
\cite{nikiforov, lorente} which again requires the use of the properties of the Laguerre polynomials.

In the present paper, we will show that the properties of the generalized Laguerre polynomials
can be obtained as a by product of the factorization of the radial
equation for the isotropic harmonic oscillator in arbitrary dimensions.
We also show that the same procedure can be applied to obtain
bound energy eigenstates of the Morse oscillator as well.

The paper is organized as follows. In Section II we set our notation by defining the ladder operators that factorize the radial part of the
 Schr\"{o}dinger equation for the isotropic harmonic oscillator, following the method of Infeld-Hull \cite{Infeld-Hull}.
Using these ladder operators, we also define  new type of $l$-changing ladder operators
which leave value of radial quantum number $j$ fixed. We also show how to construct further new ladder operators
that shift between any  given pair of radial eigenstates labeled by angular and radial quantum numbers $(l,j)$.

In Section III we indicate  how the properties of the generalized Laguerre polynomials can be
obtained by using the various ladder operators and the radial eigenfunctions defined  in Section II.

In Section IV we find  the energy eigenstates of the Morse oscillator in terms of the generalized
Laguerre polynomials by using the ladder operators obtained by factorizing the Morse hamiltonian.
In Section V we present our conclusions.

 \section{Isotropic Harmonic Oscillator In $d$-dimensions}
The details  of the factorization method presented here can be found in \cite{Infeld-Hull} where
the isotropic harmonic oscillator is of classified as type-C factorization.
We also refer to \cite{dong} for the factorization method applied to the radial
Schr\"odinger equation for a non-relativistic particle
with the central potential of the form $\frac{1}{2}m\omega^2r^2+\frac{\hbar^2}{2m}\frac{\alpha}{r^2}$ 
in $d$ dimensions from a different point of view.

 We shall work with the variables expressed in natural units.
The natural unit of energy for  the harmonic oscillator is $\hbar\omega$. Dividing the
Schr\"{o}dinger equation by $\hbar\omega$ and also scaling length by $r_0\equiv(\frac{\hbar}{m\omega})^{1/2}$;
momentum by $\hbar/r_0$ one has the Schr\"{o}dinger equation in terms of the natural units. 
In the natural unit system, the canonical commutation relations read  $[x_m,p_n]=i\delta_{mn}$ where $m,n=1,2,\ldots d$.

Up to a sign,  square of the momentum in coordinate representation is  given by the generalized Laplacian  operator. 
In Euclidian space $I\!\!R^d$, and  in terms of spherical  coordinates, it can be written as
\be\label{laplace-beltrami}
\Delta^{{I\!\!R}^d}
=
\frac{1}{r^{d-1}}\frac{\partial}{\partial r}r^{d-1}\frac{\partial}{\partial r}
+
\frac{1}{r^2}\Delta^{S^{d-1}}
\ee
where $\Delta^{S^{d-1}}$ is the Laplacian operator defined on the unit sphere $S^{d-1}$
with the metric induced from that of the ambient Euclidian space $I\!\!R^d$. The radial momentum, 
in accordance with the Weyl prescription of ordering of non-commuting operators, can be defined as
\be
p_r
=
\frac{1}{2}\left(\frac{1}{r}\vec{r}\cdot\vec{p}+\vec{p}\cdot\vec{r}\frac{1}{r}\right)
=
\frac{-i}{r^{(d-1)/2}}\frac{\partial }{\partial r}r^{(d-1)/2}
\ee
in the position representation.
 The radial momentum has the canonical commutator
$[r,p_r]=i$ with the radial coordinate
and one has $p_r=p_r^\dagger$ with
respect to the inner product defined in terms of the
the weight function $w(r)=r^{(d-1)}$
with $r\in [0,\infty)$. With this weight function, the inner product of two
functions $|f\rangle$ and $|g\rangle$, in the Hilbert space of
functions spanned by the square integrable energy eigenfunctions becomes
\be\label{inner-prod}
\langle f|g\rangle
=
\int^{\infty}_{0}drr^{d-1}f^*(r)g(r).
\ee
The radial term  of the Laplacian operator in (\ref{laplace-beltrami})
can be written in terms of the coordinate representation of the radial momentum as
\be
-\frac{1}{r^{d-1}}\frac{\partial}{\partial r}r^{d-1}\frac{\partial}{\partial r}
=
p_r^2+(d-1)(d-3)\frac{1}{4r^2}.
\ee
$(d-1)$ number of angular coordinates can be separated from the radial coordinate by introducing
generalized spherical harmonics, $\mathcal{Y}_l(\hat{r})$,  in $d$ dimensions  \cite{joseph1}.
Recalling
that $r^l\mathcal{Y}_l(\hat{r})$ can be expressed in terms of homogeneous irreducible monomials
of order $l$ in the Cartesian coordinates $x_m$. Hence,  they satisfy
$\Delta^{{I\!\!R}^d}r^l\mathcal{Y}_l(\hat{r})=0$ and therefore using (\ref{laplace-beltrami}) one has
\be
\Delta^{S^{d-1}}\mathcal{Y}_l(\hat{r})
=
-l(l+d-2)\mathcal{Y}_l(\hat{r}).
\ee
As a result, one obtains the set of effective  hamiltonians  $h_l$, depending on the radial coordinate
and also on the separation parameters $l$ and $ d$. $h_l$
can be written as
\be\label{dho-eff-hamiltonian}
h_l
=
\frac{1}{2}
\left\{p_r^2 +
\left[\left(l+\tfrac{1}{2}(d-2)\right)^2-\tfrac{1}{4}\right]\frac{1}{r^2}
+
r^2\right\}.
\ee
This form of $h_l$
implies that  the eigenvalue equation
$h_l|l,j\rangle =\varepsilon_j(l) |l,j\rangle$ admits factorization since it is of the form the sum of the squares of two operators up to a constant
remainder.
Note also that the radial Coulomb-type potential of the form $\frac{1}{r}$ also allows the corresponding
effective radial hamiltonian to be written
as the of product  of two
hermitian conjugate operators up to a constant remainder.

The first label of  radial eigenkets $|l,j\rangle$ refers to
the angular momentum quantum number, whereas the second
one $j$ refers to the radial quantum number.
For convenience,
the energy eigenvalues corresponding to the radial eigenfunctions $R^{(l)}_0(r)=:\langle r|l,0\rangle$ are  denoted by
$\varepsilon_j(l)$,
all of which depend
 on the parameters $l,j$ and $d$.

The ladder operators, which are
the
conjugates of each other with  respect to the inner product (\ref{inner-prod}) defined with respect to the
weight function $w(r)=r^{(d-1)}$, are of the form
\begin{eqnarray}
D_l
&=&
\frac{1}{\sqrt{2}}\left(+ip_r-\left(2l+d-1\right)\frac{1}{2r}+r\right),
\label{step-down-def0}\\
D^\dagger_l
&=&
\frac{1}{\sqrt{2}}\left(-ip_r-\left(2l+d-1\right)\frac{1}{2r}+r\right).
\label{step-up-def0}
\end{eqnarray}
By using the same symbols  for the position representations for the ladder operators, the position representations
of these operators can be written  as
\begin{eqnarray}
D_l
&=&
+\tfrac{1}{\sqrt{2}}r^{l}e^{-r^2/2}\frac{d}{dr}e^{r^2/2}r^{-l},
\label{step-down-def}\\
D^\dagger_l
&=&
-\tfrac{1}{\sqrt{2}}r^{-(l+d-1)}e^{r^2/2}\frac{d}{dr}e^{-r^2/2}r^{(l+d-1)}.
\label{step-up-def}
\end{eqnarray}
The particular  forms  of the
position representations (\ref{step-down-def}) and (\ref{step-up-def})
for the first order ladder operators can be obtained from (\ref{step-down-def0}) and (\ref{step-up-def0}) respectively by
introducing appropriate integrating factors for the multiplicative terms of the ladder operators
and will be shown to be very useful for
the discussions  below.

For $l\mapsto0$ and $d\mapsto1$, the expressions in (\ref{step-down-def}) and (\ref{step-up-def})
become the position representations for the ladder operators of the one dimensional harmonic oscillator, namely
$a$ and $a^\dagger$ respectively.

The choice of
the
ladder operators having the same algebraic structure
 is  not unique. However, the  eligible ladder operators
 can be obtained from the
requirements: (i)
The key functions which are annihilated by $D_l$ should be normalizable, that is, $\langle l,0|l,0\rangle=1$,
(ii) $\varepsilon_j(l)<\varepsilon_{j+1}(l)$ for all $l,j=0,1,2,\ldots$
which  ensures the normalization of the radial eigenkets $|l,j\rangle$, that is, $\langle l,j|l,j\rangle=1$ for $j>0$.
In terms of the  ladder operators above, the hamiltonian can be written as
\begin{equation}
h_l
=
D^\dagger_{l}D_{l}+\varepsilon{(l)}
\end{equation}
where $\varepsilon_0(l)=l+d/2$ is the ground state energy eigenvalue corresponding to the key function $R^{(l)}_0$
provided that $D_lR^{(l)}_0=0$. The spectra of the sequence of the hamiltonians $h_{(l+j)}$
with $l,j=0,1,2\ldots$ follow from the algebra of the ladder operators (see Fig. 1).
From the definitions of the ladder operators, one finds the recurrence relations
\be\label{SI1}
D^\dagger_{(l+j+1)}D_{(l+j+1)}+2 = D_{(l+j)}D^\dagger_{(l+j)}.
\ee
with $l,j=0,1,2\ldots$. In terms  of the sequence of the hamiltonians $h_{(l+j)}$ which are naturally defined as
\be
h_{(l+j)}
=
D^\dagger_{(l+j)}D_{(l+j)}+\varepsilon_j{(l)}.
\ee
The recurrence relation (\ref{SI1}) for $j=0$ can be written as
\be\label{lower-SIP}
h_lD^\dagger_{l}
=
D^\dagger_lh_{(l+1)}.
\ee
Iteration of this result in the radial quantum number $j$ gives
\be\label{recur-hamil-ladder}
h_l D^\dagger_l D^\dagger_{(l+1)}\cdots D^\dagger_{(l+j-1)}
=
D^\dagger_l D^\dagger_{(l+1)}\cdots D^\dagger_{(l+j-1)}h_{(l+j)}.
\ee
The recurrence relation (\ref{recur-hamil-ladder}) implies that
the  ground state energy of  $h_{(l+j)}$ corresponds to the $j^{th}$ excited level of $h_l$
which is  given by $\varepsilon_j(l)
=
l+2j+d/2$.
Thus, the radial energy eigenfunction  corresponding to the $j^{th}$  excited eigenstate
of the hamiltonian $h^{}_{l}$ is given by
\be\label{radial-def-l-j}
|l,j\rangle
\propto
D^\dagger_{l} D^\dagger_{(l+1)}
D^\dagger_{(l+2)}\cdots  D^\dagger_{(l+j-1)}|l+j,0\rangle
\ee
up to an appropriate normalization constant to be found below.
 The radial key eigenfunctions are annihilated by the ladder operator
\be\label{annihilation}
D_{(l+j)}|l+j,0\rangle
=
0.
\ee
This yields the normalized key radial eigenfunctions as
\be
\langle r|l+j,0\rangle
=
R^{(l+j)}_0(r)
=
\left[\tfrac{1}{2}\Gamma(l+j+d/2)\right]^{-1/2}
r^{(l+j)}e^{-r^2/2}
\ee
by inspecting  the position representation of the ladder operator in (\ref{step-down-def}).
It follows from both the  definition  of the key functions and of the ladder operators in harmony that
\be
D_{(l+j)}|l+j,0\rangle
=
rD_{(l+j-1)}r^{-1}|l+j,0\rangle=0
\ee
and as a result one finds $|l+j,0\rangle\propto r^{\mp}|l+j\mp1,0\rangle$. Therefore, the operators $r^{\mp1}$ can be identified as
the ladder operators which shift between the successive key functions.
Furthermore, the definition of the  operator $D^\dagger_{(l+j)}$ in (\ref{step-up-def}) allows
the ladder operators $D^\dagger_{(l+j)}$ to be written in terms of  $D^\dagger_{0}$ as
\be\label{step-down-recur}
D^\dagger_{(l+j)}
=
r^{-(l+j)}D^\dagger_{0} r^{(l+j)}.
\ee
The relation in (\ref{step-down-recur}) and its conjugate are very practical  in the calculations below.
Using (\ref{step-down-recur}), it is possible to rewrite all the excited radial eigenstates of the hamiltonians $h_l$ as
\begin{eqnarray}
|l,j\rangle
&\propto&
D^\dagger_{l}D^\dagger_{(l+1)}
D^\dagger_{(l+2)}\cdots  D^\dagger_{(l+j-1)}|l+j,0\rangle,
\label{recur00}\\
R^{(l)}_j(r)
&\propto&
r^{-(l+d-2)}e^{r^2/2}\left(\frac{-1}{\sqrt{2}r}\frac{d}{dr}\right)^j
r^{[2(l+j-1)+d]}e^{-r^2}\label{radial-eigenstate-form1}
\end{eqnarray}
up to a normalization constant. The normalization of the eigenfunctions $R^{(l)}_j(r) $ follows
from the normalization of the eigenfunctions $R^{(l+j)}_0(r) $. Using (\ref{SI1}) and  by induction   one finds
\begin{eqnarray}
\langle l,j|l,j\rangle
&=&
\langle l+j,0 |D_{(l+j-1)}D_{(l+j-2)}\cdots D_lD^\dagger_{l}
\cdots D^\dagger_{(l+j-1)}
|l+j,0\rangle
\\
&=&
\prod^{k=j}_{k=1}(\varepsilon_{k}(l)-\varepsilon_0(l))\langle l+j,0|l+j,0\rangle
\\
&=& 2^{j}j!\langle l+j,0|l+j,0\rangle.
\end{eqnarray}
Therefore, the normalized radial eigenfunctions can be written in a convenient form as
\be\label{luck01}
R^{(l)}_j(r)
=
\left[2^{(j-1)}j!\Gamma(l+j+d/2)\right]^{-1/2}
e^{r^2/2} r^{-(l+d-2)} \left(-\sqrt{2}\frac{d\phantom{a1}}{d(r^2)}\right)^j
r^{2(l+d/2-1+j)}e^{-r^2}.
\ee
The expression
(\ref{luck01}) is a generalized Rodrigues-type formula  for the radial eigenfunctions
and it is easy to write it in terms of the generalized Laguerre polynomials  in the variable $r^2$
as
\be
R^{(l)}_j(r)
=
(-1)^j
\left(\frac{2\Gamma(j+1)}
{\Gamma(l+j+d/2)}\right)^{1/2}e^{-r^2/2}r^{l}L^{(l+d/2-1)}_{j}(r^2),\label{main1}
\ee
for $j\geq1$ values of the radial quantum number.
In the next section we will show that the functions $L^{(l+d/2-1)}_{j}(z)$ defined above
do indeed satisfy the generalized Laguerre differential equation
in the variable $z=r^2$ by using the recurrence relations provided by the ladder operators, cf. Eqns. 
(\ref{recur-mine2})-(\ref{laguerre-de}) below.  Note that the radial key eigenfunctions themselves are not generalized
Laguerre polynomials and that  the minimum value of $l+d/2-1$ is $-1/2$ which corresponds to
values of $l=0$ and $d=1$. The Laguerre polynomials, namely, $L^{\mu=0}_j$'s, occur only  for $d=2$ and zero angular momentum
radial energy eigenstates. The expressions (\ref{luck01}) and (\ref{main1}) hint
at  the possibility that the ladder operators can be used to obtain the properties of the generalized 
Laguerre polynomials which are usually obtained by other means
\cite{nikiforov}.The functions $e^{-z/2}z^\mu L^{\mu}_{j}(z)$ are sometimes called the 
\emph{generalized} Laguerre functions \cite{lorente}. In Section IV, we will also show that
for $z=2e^{-x}$ these functions correspond to the bound  eigenstates of the Morse oscillator.
In the study of the coherent states defined in
terms of the ladder operators, a hamiltonian whose eigenfunctions are \emph{generalized} Laguerre functions
were  constructed in \cite{jellal}. In a more general scheme, it is also possible to
define ladder operators for the classical polynomials starting from the corresponding generalized Rodrigues formula
and the recurrence  relations \cite{lorente,cotfas}.
\begin{figure}[t]
\setlength{\unitlength}{1mm}
\begin{picture}(150,70)(0,0)

\put(10,10){\vector(1,0){120}}

\put(10,10){\vector(0,1){60}}
\put(8,72){$\varepsilon_j(l)$}
\multiput(20,10)(20,0){6}{\line(0,1){.9}}
\put(20,7){$l$}
\put(38,7){$l+1$}
\put(58,7){$l+2$}
\put(78,7){$l+3$}
\put(98,7){$l+4$}
\put(118,7){$l+5$}

\multiput(10,20)(0,7.5){7}{\line(1,0){1}}

\put(28,65){$D_{l}$}\put(25,58){$D^\dagger_{l}$}
\put(50,58){$D_{(l+1)}$}\put(43,51){$D^\dagger_{(l+1)}$}
\put(70,50){$D_{(l+2)}$}\put(63,44){$D^\dagger_{(l+2)}$}

\put(17,20){$|l,0\rangle$}\put(-1,19){$\varepsilon_0(l)$}
\put(17,35){$|l,1\rangle$}\put(-4,34){$\varepsilon_0(l)+2$}

\put(24,39){\vector(2,1){8}}\put(22,43){$\mathcal{D}^\dagger_{(l,1)}$}
\put(32,42){\vector(-2,-1){8}}\put(28,37){$\mathcal{D}_{(l,1)}$}

\put(17,50){$|l,2\rangle$}\put(-4,49){$\varepsilon_0(l)+4$}

\put(24,65){\vector(2,-1){8}}
\put(32,60){\vector(-2,1){8}}

\put(46,58){\vector(2,-1){8}}
\put(54,53){\vector(-2,1){8}}

\put(65,51){\vector(2,-1){8}}
\put(73,46){\vector(-2,1){8}}

\put(85,46){\vector(2,1){6}}\put(86,48){$r$}
\put(91,48){\vector(-2,-1){6}}\put(87,43){$r^{-1}$}

\put(17,65){$|l,3\rangle$}\put(-4,64){$\varepsilon_0(l)+6$}

\put(20,70){$\vdots$}
\put(58,70){$\vdots$}
\put(98,70){$\vdots$}

\put(33,27.5){$|l+1,0\rangle$}
\put(33,42.5){$|l+1,1\rangle$}
\put(33,57.5){$|l+1,2\rangle$}
\put(38,65){$\vdots$}

\put(52,35){$|l+2,0\rangle$}
\put(52,50){$|l+2,1\rangle$}
\put(52,65){$|l+2,2\rangle$}

\put(78,65){$\vdots$}
\put(118,65){$\vdots$}

\put(72,42.5){$|l+3,0\rangle$}
\put(72,57.5){$|l+3,1\rangle$}

\put(92,50){$|l+4,0\rangle$}
\put(92,65){$|l+4,1\rangle$}

\put(112,57.5){$|l+5,0\rangle$}

\end{picture}
\caption{The lattice of the energy levels for the sequence of
hamiltonians $h_{(l+j)}$. The column of the equally spaced eigenkets above 
the ground level $|l+j,0\rangle$ for each $j=0,1,2\ldots$
constitute the  spectrum of the hamiltonian $h_{(l+j)}$.  The energy
difference between successive key eigenkets is one unit (of
$\hbar\omega$) whereas  the energy difference of the
two successive eigenkets of the hamiltonian $h_{(l+j)}$ is two units. The actions of the
ladder operators on the eigenkets are indicated by the arrows. }
\end{figure}
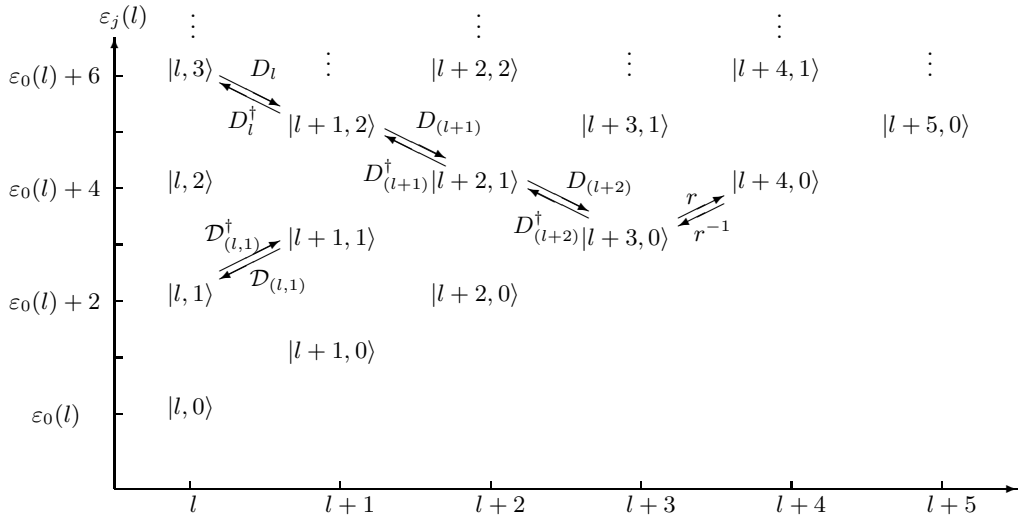

It is possible to construct  a new type of ladder operators, namely $l$-changing ladder operators,
which  shift between the radial eigenfunctions with a fixed value of the radial quantum number $j$.
For this purpose,  first  recall that $|l+j\mp1,0\rangle\propto r^{\mp1}|l+j,0\rangle$  and
therefore, using this property, the relation (\ref{radial-def-l-j})
can be written as
\begin{eqnarray}
|l+1,j\rangle
&\propto&
D^\dagger_{(l+1)} D^\dagger_{(l+2)}
D^\dagger_{(l+3)}\cdots  D^\dagger_{(l+j)}|l+j+1,0\rangle
\nonumber\\
&=&
r^{-1}D^\dagger_{l} D^\dagger_{(l+1)}
D^\dagger_{(l+2)}\cdots   D^\dagger_{(l+j-1)} r^2 |l+j,0\rangle\label{pass1}
\end{eqnarray}
up to a normalization constant. By commuting $r^2$ to the left of the product of the ladder operators, 
it is possible to rewrite the right hand side back  in terms of $|l,j\rangle$ again.  To this end, one 
needs the commutator identity
\be\label{last-cr}
\left[
 D^\dagger_{l} D^\dagger_{(l+1)}
D^\dagger_{(l+2)}\cdots   D^\dagger_{(l+j-2)} D^\dagger_{(l+j-1)}, r^2
\right]
=
-
\sqrt{2}j r D^\dagger_{(l+1)}
D^\dagger_{(l+2)}\cdots   D^\dagger_{(l+j-2)} D^\dagger_{(l+j-1)}.
\ee
The commutator (\ref{last-cr}) can  be found by using (\ref{step-down-recur}) and the canonical 
commutation relation $[r,p_r]=i$. Using this result (\ref{pass1}) takes the form
\be
|l+1,j\rangle
\propto
-\sqrt{2}j|l+1,j-1\rangle+r
 |l,j\rangle
\ee
up to an overall  constant. By using  (\ref{luck01}), this can be written in terms of the ladder operators as
\be \label{ww1}
|l+1,j\rangle
\propto
e^{r^2}D_{l}e^{-r^2}
 |l,j\rangle.
\ee
again up to an overall constant. The operator on the right hand side of (\ref{ww1}) evidently
defines an $l$-changing  ladder operator which will be denoted by
$\mathcal{D}^\dagger_{(l,j)}$. Using the above notation, the action of this new ladder
operator on the radial eigenkets can be written as
\be\label{l-ladder-up}
\mathcal{D}^\dagger_{(l,j)}|l,j\rangle
\propto
|l+1,j\rangle.
\ee
The conjugate of this operator  defines yet another $l$-changing ladder operator,
$\mathcal{D}_{(l,j)}$,  which can be  simply written as
\be\label{l-ladder-down}
|l+1,j\rangle
\propto
\mathcal{D}_{(l,j)}|l,j\rangle
=
e^{-r^2}D^\dagger_{l}e^{r^2}|l,j\rangle
\ee
up to a constant  depending on both of the quantum numbers.  These new $l$-changing  
ladder operators $\mathcal{D}^\dagger_{(l,j)}$ and $\mathcal{D}_{(l,j)}$
depend on both of the  quantum numbers
$l$ and $j$
and also on the dimension $d$ which is not indicated for convenience of the notation.
Note here that, it is possible to define
an operator composed of appropriate combinations of the ladder operators defined so far,
namely,  $D^\dagger_{(l+j)}$, $D_{(l+j)}$, $\mathcal{D}^\dagger_{(l,j)}$,
$\mathcal{D}_{(l,j)}$ and $r^{\mp1}$,
that shifts between any given pair of the eigenkets $|l,j\rangle$ and $|l',j'\rangle$ illustrated in Fig. 1.
For instance, the combination  $D^\dagger_{l}\mathcal{D}^\dagger_{(l,j)}$  of the ladder operators
acting on the eigenstate $|l,j\rangle$ gives
\be
D^\dagger_{l}\mathcal{D}^\dagger_{(l,j)}|l,j\rangle
\propto
|l,j+1\rangle
\ee
up to a normalization constant. Therefore, it is possible to identify  $D^\dagger_{l}\mathcal{D}^\dagger_{(l,j)}$ and its conjugate, namely
$\mathcal{D}_{(l,j)}D_{l}$,
as   $j$-changing  ladder operators.

It is worth  to emphasize at this point  that (i)The ladder operators above
can be used to  construct  new ladder operators that shift
between any given states $|l,j\rangle$ and $|l',j'\rangle$ and  $l$-changing   and $j$-changing operators defined above
are examples of these  new  of ladder operators which shift between $|l\mp1,j\rangle$ and $|l,j\mp1\rangle$ respectively,
(ii) Nowhere in the above construction  the properties of the generalized Laguerre polynomials have
been used. Furthermore,  various recurrence relations for  the generalized
Laguerre polynomials can be derived using any of the ladder operators defined above. From this point of view,
the presentation above can be considered to unify the method of solving eigenvalue problem and obtaining the recurrence relations of  the related
special function as a by product. In contrast to the other approaches, the ladder operators are
constructed by invoking the properties of related special functions \cite{liu, cardoso} and
the  matrix elements for  any of the ladder operators defined above can be calculated using the  
mathematical tools constructed above without employing any of the properties of the Laguerre polynomials as  
additional mathematical property of the solutions. For example,  integrals involving the generalized Laguerre polynomials can be calculated by using
the normalization of the radial eigenfunctions and the expressions of the radial eigenfunctions in terms  of the ladder
operators, in particular, the conventional  normalization of the generalized Laguerre polynomials can be found from the
normalization of the radial eigenfunctions.

As will be shown in the next section, the expressions involving the  ladder operators and the energy eigenfunctions 
can be translated into the expressions involving the generalized Laguerre polynomials. Since it is possible to construct a number of 
different recurrence relations for the radial eigenfunctions  by defining an appropriate composition 
of the ladder operators defined above, the factorization method also provides
a new way of deriving new recurrence relations for the generalized Laguerre polynomials as well.

\section{Properties of the Generalized Laguerre Polynomials }
For the definitions and
the
conventions of the Laguerre polynomials adopted here we refer to \cite{guo,nikiforov}.
First, we  note that the standard normalization of the generalized Laguerre polynomials follows easily from
the normalization of the radial eigenfunctions,
all of
which can be turned into Gamma function-type integrals.
The radial eigenfunctions which are normalized to unity are given explicitly in (\ref{luck01}) and they
can be expressed  in terms of the Laguerre polynomials using
(\ref{main1}). Therefore, the normalization of the radial eigenfunctions $|l,j\rangle$ can be expressed in terms of the
Laguerre polynomials as
\begin{eqnarray*}
\langle l,j |l,j\rangle
=
1
&=&
\int^{\infty}_{0}drr^{(d-1)}|R^{(l)}_{j}(r)|^2
\\
&=&
\frac{\Gamma(j+1)}{\Gamma(l+d/2+j)}
\int^{\infty}_{0}dz e^{-z}z^{l+d/2-1}|L^{(l+d/2-1)}_{j}(z)|^2
\end{eqnarray*}
where the variable $r^2=z$ has been introduced in the second line. From this expression
it follows that
 \be\label{norm-laguerre}
\int^{\infty}_{0}dze^{-z}z^{\mu}|L^{\mu}_{j}(z)|^2
=
\frac{\Gamma(\mu+j+1)}{\Gamma(j+1)}
\ee
with $\mu=l+d/2-1$. This standard normalization can also
be obtained by means
of the Rodrigues formula for the Laguerre polynomials.

Second, we illustrate that
any expression for the radial eigenfunctions involving the ladder operators
can be translated into corresponding  recurrence relation for the Laguerre
polynomials and in particular we derive the Laguerre differential equation as follows.
As the simplest example, consider the expression (\ref{recur00}). By using the
relations among the radial eigenstates $R^{(l)}_{j}$ and $R^{(l+1)}_{(j-1)}$, namely,
\be\label{rc1}
|l,j\rangle\propto D^\dagger_l |l+1,j-1\rangle
=
r^{-l}D^\dagger_0 r^{l} |l+1,j-1\rangle
\ee
it is possible to derive  corresponding recurrence relation for the generalized Laguerre polynomials.
By the change of variable $r^2=z$ and using the definition (\ref{main1}), Eqn. (\ref{rc1}) becomes
\be\label{recur-mine2}
L^{(l+d/2-1)}_{j}(z)
=
\frac{1}{j}z^{-(l+d/2-1)}e^{z}\frac{d}{dz}e^{-z}z^{(l+d/2)}
L^{(l+d/2)}_{(j-1)}(z).
\ee
Similarly, by using the normalized ladder operator $D_l$, the recurrence relation
\be
|l+1,j-1\rangle =D_l|l,j\rangle,
\ee
in terms of Laguerre polynomials, becomes
\be\label{recur-mine-add2}
L^{(l+d/2-1)}_{j}(z)
=
-\frac{d}{dz}L^{(l+d/2)}_{j-1}(z)
\ee
where $z=r^2$. The recurrence relations  (\ref{recur-mine2}) and (\ref{recur-mine-add2}) can be used to obtain the second order
differential equation that $L^{(l-1+d/2)}_{j}(z)$ satisfy. By defining  $\mu=l+d/2-1$ for convenience and using the relations
(\ref{recur-mine2}) and (\ref{recur-mine-add2}) one finds
\be\label{laguerre-de}
z\frac{d^2}{dz^2}L^{\mu}_{j}(z)+(\mu+1-z)\frac{d}{dz}L^{\mu}_{j}(z)+jL^{\mu}_{j}(z)=0
\ee
which justifies, 
a posteriori,
that  the $L^{\mu}_{j}$ introduced  in (\ref{main1}) corresponds to the  generalized Laguerre polynomials.

For $d=1$ the quantum number corresponding to the angular momentum can take 
the value zero only while the radial quantum number becomes the principle quantum number for the 1-$d$ harmonic oscillator
and therefore for  the values $d=1$, $l=0$ the radial eigenfunctions become  the energy eigenfunctions of the 1-$d$ harmonic oscillator.
By using the  expression (\ref{radial-eigenstate-form1}) for the radial eigenfunctions $R^{(l)}_j$
and the key functions $R^{(l)}_0$ together with  the definition of the Hermite polynomials, one has the relation
\be\label{Hermite-Laguerre}
H_{2j}(r)
=
(-1)^j2^jj!L^{-1/2}_j(r^2)
\ee
where $H_j(z)$ is the Hermite polynomial of order $j$ and $L^{\mu}_j(z)$ is a polynomial of order $j$ in the 
variable $z$ respectively \cite{guo}. Using this identification, it is easy to  check the validity of the 
recursion relation (\ref{recur-mine2}). This can be done by putting $d=1$ and $l=0$
 in (\ref{recur-mine2}). Thus,  by the change of the variable $r^2=z$, one finds
\be
L^{-1/2}_{j}(z)
=
-z^{-1/2}e^{z}\frac{d}{dz}e^{-z}z^{1/2}
L^{1/2}_{(j-1)}(z).\label{recur-form-2}
\ee
Using the relation (\ref{Hermite-Laguerre}), and the recursion relation between the  Hermite polynomials, namely,
  \be
 H_{(2j+1)}(r)
 =
 -\frac{1}{\sqrt{2}}e^{r^2}\frac{d}{dr}e^{-r^2} H_{2j}(r)
 \ee
 which already follows  from the construction above, one finds that
\be
L^{1/2}_{j}(r^2)
=
\frac{(-1)^j}{2^jj!r}H_{(2j+1)}(r).
\ee
Returning back to the  recurrence relation (\ref{recur-mine2}), it
can be can be put into a more useful form
\be
z^{-\mu}e^{-z}L^{\mu}_{j}(z)
=
\frac{1}{j}\frac{d}{dz}e^{-z}z^{(\mu+1)}
L^{(\mu+1)}_{(j-1)}(z)
\ee
with $\mu=l+d/2-1$ for convenience. This form of the recurrence relation shows that $\frac{d}{dz}$ is
the ladder operator between the functions $e^{-z}z^{\mu}L^{\mu}_{j}(z)$ and  can be iterated $k$ times
to obtain
\be\label{general-recur2}
L^{\mu}_{j}(z)
=
\frac{(j-k)!}{k!}z^{-\mu}
e^{z}
\frac{d^k}{dz^k}e^{-z}
z^{(\mu+k)}L^{(\mu+k)}_{(j-k)}(z).
\ee
In terms of the radial coordinate and in terms of the radial eigenfunctions,
this recurrence relation corresponds to the coordinate  representation of the expression
\begin{eqnarray}
|l,j\rangle
&\propto&
D^\dagger_{l} D^\dagger_{(l+1)}
D^\dagger_{(l+2)}\cdots D^\dagger_{(l+k-2)} D^\dagger_{(l+k-1)}|l+k,j-k\rangle.
\label{general-recur1}
\end{eqnarray}
For $k=j$, this yields the Rodrigues formula for the Laguerre polynomials rather
then a recurrence relation among them. Similarly, the conjugate of the relation (\ref{general-recur1}), namely,
\be\label{pre-tr1}
|l+k,j-k\rangle
\propto
D_{(l+k-1)}D_{(l+k-2)}\cdots  D_{(l+2)}D_{(l+1)}D_{l} |l,j\rangle
\ee
can be translated into the  recurrence relation for the generalized Laguerre polynomials as
\be
L^{(\mu+k)}_{(j-k)}(z)
=
(-1)^k\frac{d^k}{dz^k}L^{\mu}_{j}(z).
\ee
The expressions for the radial eigenfunctions involving $l$-changing operators  $\mathcal{D}^\dagger_{(l,j)}$
and $\mathcal{D}_{(l,j)}$ can also be translated into a recurrence relation for the generalized Laguerre
polynomials. For example, using normalized ladder operators, the  relation (\ref{l-ladder-down})
can used to derive  recurrence relations for the generalized  Laguerre polynomials as
\be
L^{(\mu-1)}_j(z)
=
z^{-\mu+1}\frac{d}{dz}
z^{\mu}L^{\mu}_j(z),
\ee
where $\mu=l+d/2-1$ and $z=r^2$ as before. Evidently, it is possible to 
derive further recurrence relations for the Laguerre polynomials from the 
expressions involving the ladder operators acting on the radial eigenfunctions. 
As the above typical examples illustrate, the normalization and recurrence properties of
the Laguerre polynomials are built into the factorization method by construction.

\section{Morse Oscillator}
The above  choice and the special way of writing the
ladder operators $D_{(l+j)}$ and $D^\dagger_{(l+j)}$ for the harmonic oscillator hamiltonian
make it possible to obtain
the useful properties of the generalized Laguerre
polynomials. We will illustrate that the above approach also provides a framework
for expressing energy eigenstates of factorizable (or supersymmetric) hamiltonians
in terms of  related special function.

In this section we shall obtain the bound energy eigenstates of the Morse
oscillator following the same approach as the harmonic oscillator.
The Morse potential, which is important in the study of molecular vibrations, is given by
\be
V(x)
=
 V_0(e^{-2\beta x}-2e^{-\beta x})
 \ee
with the parameters $V_0>0$ and $\beta$ having dimension inverse length with $-\infty<x<\infty$
\cite{Infeld-Hull,morse,il-cooper}. It is also possible to transform the Morse potential problem into the two dimensional
harmonic oscillator by an appropriate change of variables \cite{dayi-duru}.  The finite number energy 
eigenstates for  the bound states of the Morse potential can be expressed  in terms of the generalized 
Laguerre polynonmials by constructing appropriate ladder operators as follows. The natural units of length, 
momentum and energy of the Morse hamiltonian can be taken as  $1/\beta$, $\hbar\beta$ and $\frac{(\hbar\beta)^2}{2m}$ 
respectively. By defining the parameter $\lambda=\frac{2V_0m}{(\hbar\beta)^2}$, and shifting the coordinate by $-\frac{1}{\beta}\ln\lambda$, the
Schr\"odinger equation for the Morse potential can be written as
\be
\left[p^2
+
\left(e^{-x}-\lambda \right)^{2}\right]\psi
=
\left\{\lambda^2-|\varepsilon|\right\}\psi
=
\varepsilon'\psi
\ee
which is convenient  for our purposes. Therefore, the  suitable ladder operators that factorize the 
Morse hamiltonian are of the form
\begin{eqnarray}
D_n
&=&
+ip-(e^{-x}-n),
\label{morse-l-down0}\\
D^\dagger_n
&=&
-ip-(e^{-x}-n).\label{morse-l-up0}
\end{eqnarray}
As in the case of the harmonic oscillator, by using the same symbols for the operators,
the coordinate representation of the conjugate ladder operators can be written conveniently as
\begin{eqnarray}
D_n
&=&
+e^{-nx-e^{-x}}
\frac{d}{dx}e^{nx+e^{-x}},
\label{morse-l-down1}\\
D^\dagger_n
&=&
-e^{nx+e^{-x}}
\frac{d}{dx}e^{-nx-e^{-x}},\label{morse-l-up1}
\end{eqnarray}
by multiplying from the right with  suitable integrating factors for the multiplicative terms
and inverses of the integrating factors from the left of the derivative. The recurrence relation 
for the ladder operators can be found to be
\be\label{recur-morse}
D_{(n+1)}D^\dagger_{(n+1)}
=
D^\dagger_{n}D_{n}+2n+1,\qquad n=1,2,3\ldots
\ee
by using the canonical commutation relation $[x,p]=i$. It is convenient to work with the hamiltonian $h_n$ 
expressed in terms of the ladder operators (\ref{morse-l-down0}) and (\ref{morse-l-up0}),
is given by
\be\label{hamilton-morse}
h_n
=
p^2+\left[e^{-x}-(n+\tfrac{1}{2})\right]^2
=
D_n^\dagger D_n+(n+\tfrac{1}{4}).
\ee
The single key eigenket, which corresponds to the ground state, will be denoted by  $|n\rangle$. The key
eigenket is annihilated by  the lowering operator. Thus $D_n|n\rangle=0$ yields the normalized eigenket
\be\label{morse-key}
\psi_n(x)
=
\frac{2^ne^{-nx-e^{-x}}}{[\Gamma(2n)]^{1/2}}
\ee
corresponding to the ground state energy $\varepsilon'_n=n+\tfrac{1}{4}$.  Using the recurrence relation (\ref{recur-morse}) 
and the definition of the Morse Hamiltonian (\ref{hamilton-morse}), one finds
\be\label{morse-recur2}
h_{(n+j)}
D^\dagger_{(n+j)}\cdots D^\dagger_{(n+1)}
=
D^\dagger_{(n+j)}\cdots D^\dagger_{(n+1)}\left\{h_n
+
\sum^{j}_{k=1}[2(n+k)+1]\right\}
\ee
which determines all the excited states and the corresponding eigenvalues. In turn, by
using the key functions (\ref{morse-key}) and the expression (\ref{morse-recur2}) and the position representation of the
raising ladder operator (\ref{morse-l-up1}), the excited states can be written as
\be\label{morse-luck1}
\psi_{(n+j)}(x)
\propto
\langle x|D^\dagger_{(n+j)}D^\dagger_{(n+j-1)}\cdots D^\dagger_{(n+1)}|n\rangle
=
(-1)^je^{(n+j+1)x+e^{-x}}\left(e^{-x}\frac{d}{dx}\right)^j
e^{-(2n+1)x-2e^{-x}}
\ee
with the corresponding energy $\varepsilon'_{(n+j)}=\varepsilon'_{n}+\sum^{j}_{k=1}[2(n+k)+1]$.
The expression (\ref{morse-luck1}) suggests that it is convenient to change to the variable $z=2e^{-x}$. 
Therefore, all the excited energy eigenstates can be written as
\be\label{morse-luck2}
\psi_{(n+j)}(x)
\propto
z^{-(n+j+1)}e^{z/2}\left(\frac{d\phantom{aa}}{d(1/z)}\right)^j
z^{(2n+1)}e^{-z}
\ee
in the new variable $z$.
Note  that with a suitable change of variables, the expression for excited states
in terms of ladder operators (\ref{morse-luck2}) resembles to  the corresponding expression for
the linear oscillator, where the excited states are obtained by repeated application of a single raising
ladder operator on the ground state. Using the identity
\be
\frac{d^j}{dz^j}f(z)
=
(-1)^j\rho^{(j+1)}\frac{d^j}{d\rho^j}\rho^{(j-1)}f(\tfrac{1}{\rho})
\ee
where $z=1/\rho$ and for any continuous function $f(z)$, it is possible to rewrite the expression 
(\ref{morse-luck2}) in terms of the generalized Laguerre polynomials
as
\be\label{morse-eigenkets}
\psi_{(n+j)}(x)
=
\left(\frac{\Gamma(j+1)}{\Gamma(2n+j+1)}\right)^{1/2}
z^{n}e^{-z/2}L^{2n}_{j}(z)
\ee
where $z=2e^{-x}$ and normalization constants of the eigenfunctions $\psi_{(l+j)}(x)$ can be found from  (\ref{norm-laguerre}).
The eigenfunctions (\ref{morse-eigenkets}) can easily be expressed back in terms of the parameters $\lambda$ and $\beta$
of the Morse hamiltonian. In contrast to the $d$-dimensional harmonic oscillator, for the Morse  hamiltonian there is 
only one quantum number yet the eigenfunctions are given in  terms of the generalized Laguerre polynomials defined  
with two parameters.

\section{Conclusion}

The Schr\"odinger equations for isotropic harmonic oscillator and the Morse potential are studied using the factorization method of Infeld-Hull
and for both of the potentials, the associated special function is identified as the generalized Laguerre polynomials and they are
expressed in terms of appropriate independent variables which directly obtained from the factorization. In the factorization method of Infeld-Hull, 
the connection  between the set of eigenfunctions  and the special functions involved,   is usually established, or rather verified, 
by comparing the results obtained from the algebraic methods with the results obtained by other methods \cite{dong}. In this respect, 
it is worth to emphasize that in the two  cases  presented above,  the factorization is carried out without transforming  the Schr\"odinger 
equation into a standard  Hypergeometric-type differential equation by the change of the dependent and the independent  variable at the outset 
and then employing the  properties of the generalized Laguerre polynomials. In contrast, the generalized Laguerre polynomials  are obtained directly 
using the expressions  involving the ladder operators of the standard factorization method. Thus, we showed in the particular case of the 
harmonic oscillator that the feature of the Infeld-Hull factorization method which directly relates the energy  eigenstates to the relevant
special functions, that is,  to the generalized Laguerre  functions as in (\ref{luck01}) and also showed that any expression
involving the ladder operators  defined for the radial eigenfunctions in Section II can be used to define a corresponding recurrence relation 
for the generalized Laguerre polynomials with the help of  the relation (\ref{main1}).

Finally, we remind that the algebraic approach  presented for the generalized Laguerre polynomials  above
is already known to apply to  other special functions. A well-known example is the quantum mechanical angular momentum.
Two different recurrence formulae for the associated Legendre polynomials $P^{m}_{l}(\theta)$ can be obtained
from the properties of the ladder operators $L_{\mp}$ using the key functions $P^{\mp l}_{l}(\theta)e^{\mp il\phi}$
and the zero angular momentum states. By writing the coordinate representation of the ladder operators
$L_{\mp}$ in a form similar to  those of the harmonic oscillator, it is possible to find the  Rodriguez type 
formulae from both $(L_-)^{(l-m)}|l,l\rangle\propto|l,m\rangle$ and $(L_+)^m|l,0\rangle\propto|l,m\rangle$ for the $P^{m}_{l}(\theta)$'s. 
Another well-known example  is the treatment of the quantum mechanical free particle in spherical coordinates.
By omitting the $r^2$ term in the harmonic oscillator hamiltonian in three dimensions,  the factorization method
can be used to obtain the spherical Bessel and Neumann functions as well as their recurrence formulae.
The different feature of  the latter example is that the normalization of spherical Bessel and Neumann functions 
(which involve  $\delta$-function normalization) can be shown to follow  from the  properties of the key functions 
corresponding to the zero angular momentum state.

\end{document}